\def\EE#1{\times 10^{#1}}
\def\ket#1{\left|#1\right\rangle}

\documentclass[aps,prl,twocolumn]{revtex4}
\usepackage{graphicx,amsmath,amssymb,ulem}
\begin{document}
\title{Multi-resonant spinor dynamics in a Bose-Einstein condensate}

\author{C.~Klempt$^1$, O.~Topic$^1$, G.~Gebreyesus$^2$, M.~Scherer$^1$, T.~Henninger$^1$, P.~Hyllus$^2$, W.~Ertmer$^1$, L.~Santos$^2$, J.J.~Arlt$^1$}

\affiliation{$^1$ Institut f\"ur Quantenoptik, Leibniz Universit\"at Hannover, Welfengarten 1, D-30167 Hannover}
\affiliation{$^2$ Institut f\"ur Theoretische Physik, Leibniz Universit\"at Hannover, Appelstra\ss{}e 2, D-30167 Hannover, Germany}

\date{\today}

\begin{abstract}
We analyze the spinor dynamics of a $^{87}$Rb $F=2$ condensate initially prepared in the $m_F=0$ 
Zeeman sublevel. We show that this dynamics, characterized by the creation of correlated atomic pairs 
in $m_F=\pm 1$, presents an intriguing multi-resonant magnetic field dependence induced by the trap inhomogeneity. This dependence 
is directly linked to the most unstable Bogoliubov spin excitations 
of the initial $m_F=0$ condensate, showing that, in general, even a 
qualitative understanding of the pair creation efficiency in a spinor condensate requires a careful consideration of the confinement.

\end{abstract}

\maketitle

Spinor Bose-Einstein condensates (BECs), consisting of atoms with non-zero spin, 
constitute an ideal scenario to investigate the interplay between internal and external degrees of freedom
in a multi-component superfluid. The competition between spin-dependent collisional interactions, Zeeman
effect, and inhomogeneous trapping results in an exciting range of fundamental phenomena.
As a result, spinor BECs have attracted  
large attention since the pioneering experiments in 
optical traps~\cite{Stenger1998} concerning their ground state 
properties~\cite{Ho1998,Ohmi1998,Koashi2000,Ciobanu2000} 
and the spinor dynamics induced by the spin-changing collisions, which allow for a coherent transfer 
between different spin components~\cite{Chang2004,Schmaljohann2004}.

Spinor BECs also provide exciting perspectives as novel sources 
of non-classical states of matter. In this sense, condensates initially prepared in the $m_F=0$ Zeeman 
sublevel are especially fascinating~\cite{Chang2004,Schmaljohann2004,Kuwamoto2004,Leslie2008}.
In that case, the creation of correlated pairs results in the growth of macroscopic populations in $m_F=\pm 1$.
This amplification process is ideally triggered by quantum spin fluctuations~\cite{KlemptNew}.
Interestingly, it closely resembles 
parametric amplification in optical parametric down conversion~\cite{Meystre2007},
opening exciting new routes for matter-wave squeezing and atomic Einstein-Podolsky-Rosen entanglement 
experiments~\cite{Pu2000,Duan2000}.

Correlated pair creation, and in general any spinor dynamics, demands a significant rate of 
spin-changing collisions. In typical experiments these collisions are  
suppressed by the quadratic
Zeeman effect (QZE) already in the presence of moderate magnetic fields~\cite{Chang2004}. 
However, the influence of the QZE at low fields is far from 
trivial~\cite{Kuwamoto2004,Kronjager2005,Zhang2005,Kronjager2006,Black2007}. 
In particular, spin-mixing can reach a pronounced
maximum for low but finite fields.
This resonance, contrary to those discussed below, has a non-linear character 
and has been explained in terms of phase matching~\cite{Kronjager2006}. 

The understanding of the magnetic-field dependence of the pair creation efficiency is hence crucial for the characterization of novel spinor-based sources of non-classical matter waves. In this Letter we show that this dependence generally presents an intriguing non-monotonous character which is crucially determined by the trap inhomogeneity and cannot be explained from the physics of homogeneous BECs~\cite{Lamacraft2007}. The pair creation efficiency is directly linked to the instability rate, which characterizes the exponential growth of the most unstable spin excitations of the initial BEC in $m_F=0$. This instability rate presents pronounced maxima and minima as a function of the applied field, which result in the striking multi-resonant magnetic field dependence of the pair creation efficiency (from $m_F = 0$ to $m_F = \pm 1$) observed in our experiments. Along with these resonances we observe characteristic magnetization patterns which depend on both the magnetic field~\cite{Lamacraft2007,Leslie2008,Sau2009} and the external confinement~\cite{Baraban2008,Sau2009}.

\begin{figure}[ht]
\centering
\includegraphics*[width=8.6cm]{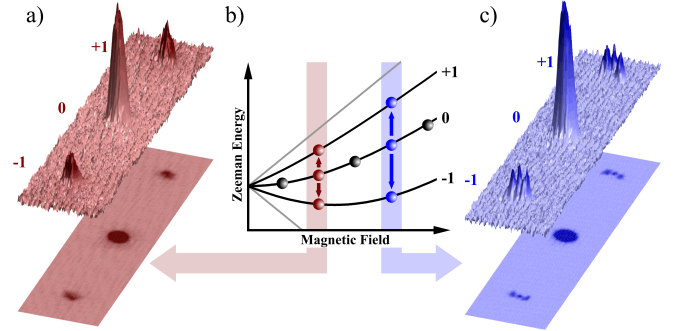}
\caption{Illustration of the resonances in the spinor dynamics
of a Bose-Einstein condensate initially in the $\ket{m_F=0}$ state. For certain magnetic fields,
the production efficiency of atoms in the $\ket{m_F=\pm 1}$ states is resonantly enhanced
due to the excess energy released by the quadratic Zeeman effect.
The column densities of all three Zeeman components in the plane of
strongest confinement are shown in (a) and (c) in both a 3D and a density
plot. Note that condensates in the $\ket{m_F=\pm 1}$ states are created in
characteristic spatial modes depending on the particular resonance.}
\label{fig1}
\end{figure}

\begin{figure}[ht]
\centering
\includegraphics*[width=8.6cm]{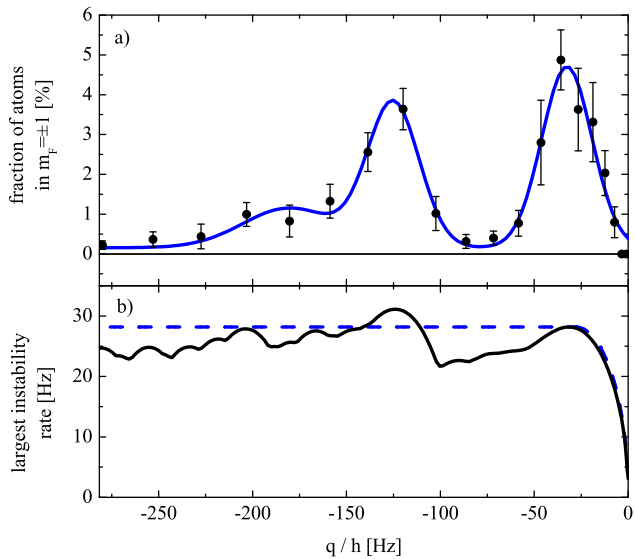}
\caption{
(a) Fraction of atoms transferred into $\ket{m_F=\pm 1}$
within $21$~ms as a function of the quadratic Zeeman energy $q$. To obtain this fraction, the number of transferred atoms per state
was divided by the sum of transferred $\ket{m_F=\pm 1}$ and condensed $\ket{m_F=0}$ atoms.
Due to strong shot-to-shot fluctuations $15$ independent realizations were averaged
at each magnetic field. The error bars indicate statistical uncertainties. The blue line
is a triple Gaussian fit to guide the eye. (b) Instability rate, given by the imaginary part of the most unstable 
spin Bogoliubov mode, corresponding to
the pair creation efficiency into $\ket{m_F=\pm 1}$ (solid line).
The maximal instability rate for an effective homogeneous case (dashed line) lacks any resonant features.}
\label{fig2}
\end{figure}

To investigate this magnetic field dependence experimentally,
we initially prepare a sample of $10^{6}$ $^{87}$Rb-atoms in the
$F=2, m_F=2$ state in a crossed beam optical dipole trap at
a wavelength of $1064$~nm (see Refs.~\cite{Klempt2008a,Klempt2008} for a more detailed description).
In the following, the $m_F=X$ states are denoted by $\ket{X}$.
The spin polarization of the atoms is maintained by a small guiding magnetic
field. The trap depth is lowered in multiple steps to
evaporatively cool the atoms from a temperature of
$2\ \mu$K to $90$~nK, resulting in a final trap depth of $150$~nK
and trapping frequencies of $(176,132,46)$~Hz. Typically a BEC of $7\EE{4}$ atoms with a thermal
fraction below $40\%$ is formed. Subsequently, the magnetic field
is raised to $78$~G and an adiabatic radio frequency passage is used to transfer the atoms
to the $\ket{0}$ state within $5$~ms. At this stage, the population of other
spin states is below the detection limit of $\approx 500$ atoms.
After the transfer, the field is lowered to $7.9$~G within $15$~ms.
During this ramp, a strong gradient of $\approx50$~G/cm is applied
to expel residual atoms in the $m_F\neq 0$ components from the trap.
Subsequent to this purification, the field is quickly lowered to a specific
value between $0.12$ and $2$~G within $3$~ms. This magnetic field strength
is calibrated by microwave hyperfine transitions and has a width of
$4$~mG due to current noise, current drift and residual gradients.
Magnetic field gradients are below $0.2$~mG/cm in all directions.
The BEC is held at the chosen magnetic field for
an adjustable time to allow for spin changing collisions. Finally, the
dipole trap is switched off to let the atomic cloud expand freely.
During time-of-flight, a strong magnetic field gradient is applied
in the vertical direction to separate the $m_F$ components.
The number of atoms in all five $m_F$ components is measured by
standard absorption imaging along the weak trap axis. A bimodal fit
to the measured density profile of the $m_F=0$ component yields the
condensed fraction and the temperature of the cloud. Typical density
profiles after time-of-flight are shown in Fig.~\ref{fig1}.

To evaluate the onset of the spinor dynamics, we restrict
our investigation to short spin evolution times and small populations in the
$\ket{\pm 1}$ states. Figure~\ref{fig2} (a) shows the magnetic field
dependence of the fraction of atoms transferred to the $\ket{\pm 1}$ states
after an evolution time of $21$ ms. The magnetic field strength $B$ is represented
by the quadratic Zeeman energy $q\propto B^2$ which is the relevant energy scale.
As expected~\cite{Schmaljohann2004}, no atoms in
the $\ket{\pm 2}$ states were detected at this time scale.
The pair-creation efficiency shows a striking multi-resonant $q$-dependence. 
We show below that this dependence
is directly linked to the interplay between Zeeman
energy, spin-changing collisions and the inhomogeneous confinement of the ensemble.

The onset of the spinor dynamics is best understood by considering the spin excitation
modes of the initial condensate~\cite{Lamacraft2007}.
Depending on the quadratic Zeeman energy $q$ and the atomic interactions, these modes may become
dynamically unstable, triggering an exponential population growth 
in $m_F\neq 0$ states, hence initiating the spinor dynamics.
An analysis based on spin excitation modes allows for a good
understanding of recent experiments in $F=1$ $^{87}$Rb condensates, including
spin-texture formation after a quench~\cite{Sadler2006,Baraban2008}, and 
- in combination with dipolar interaction - the instability of 
externally induced helices \cite{Vengalattore2008,Cherng2008,Cherng2008a}.
The initial state may be represented by a spinor wave function 
${\mathbf \Psi}_0(\vec r)=(\psi_{-2},\psi_{-1},\psi_0,\psi_1,\psi_2)^T=(0, 0, n_0(\vec r)^{1/2}, 0, 0)^T$.
The onset of the spinor dynamics (linear regime) is described by an operator
$\hat{{\bf \Psi}}(\vec r,t)=({\mathbf \Psi}_0 (\vec r)+\delta\hat{\mathbf \Psi}(\vec r,t))e^{-i\mu t}$,
where $\mu$ is the chemical potential, $\delta\hat{{\mathbf \Psi}}(\vec r,t)= (\delta\hat\psi_{-2},\delta\hat\psi_{-1},\delta\hat\psi_{0},\delta\hat\psi_{1},\delta\hat\psi_{2})^T$
describes the spin fluctuations,
and the density fulfils $n_0(\vec r)\gg \langle \delta\hat\psi_{m_F}^\dagger(\vec r)\delta\hat \psi_{m_F}(\vec r)\rangle$.
Up to second order in $\delta\hat{\mathbf \psi}_{m_F}$ we obtain a Hamiltonian for the pair creation in  
$\ket{\pm 1}$~\cite{Lamacraft2007}:
\begin{eqnarray}
\hat H&=&\int d^3{\vec r}
\left\{
\sum_{m_F=\pm 1}\delta\hat\psi_{m_F}^\dag
\left[ \hat H_{eff} +q \right ] \delta\hat\psi_{m_F}
\right\delimiter 0 \nonumber \\
&+&\left\delimiter 0
\Omega_{eff}(\vec r)
\left [
\delta\hat\psi_{1}^\dag \delta\hat\psi_{-1}^\dag+
\delta\hat\psi_{1} \delta\hat\psi_{-1}
\right ]
\right \}.
\label{eq:H}
\end{eqnarray}
Note that we may neglect the transfer into 
$\ket{\pm 2}$ during the first stages of the 
dynamics due to the much lower transfer rate from $\ket{0}$. Hence the 
Hamiltonian is valid for the description of the initial pair creation for BECs in 
both $F=1$ and $F=2$. The effective Hamiltonian for the $\ket{\pm 1}$ components is
$\hat H_{eff}=-\hbar^2\nabla^2/2m+V_{eff}(\vec r)$, with 
$V_{eff}(\vec r)\equiv V(\vec r)+(U_{0}+U_{1})n_0(\vec r)-\mu$,
where $m$ is the atomic mass, and $V(\vec r)$ is the confining potential.
The coupling coefficient $\Omega_{eff}(\vec r)=U_{1} n_0(\vec r)$
characterizes the pair creation induced by spin-changing collisions.
The interaction strengths are 
$U_0= (7g_0+10g_2+18g_4)/35$, $U_1= (-7g_0-5g_2+12g_4)/35$ for $F=2$
and $U_0=(g_0+2g_2)/3$, $U_1=(g_2-g_0)/3$ for $F=1$ condensates. They   
are related to  $g_F=4\pi\hbar^2 a_F/m$, where $a_F$ is the $s$-wave scattering 
length for the collisional channel with total spin $F$
(only even due to symmetry)~\cite{Ho1998}.  
Note that for $^{87}$Rb $U_1>0$ for $F=2$ whereas $U_1<0$ for $F=1$.

For homogeneous BECs, $V_{eff}$ and $\Omega_{eff}$ are constants,
and the spin excitations are plane waves with wavevector $\vec k$ and energy
$\xi_k (q)$ obtained after diagonalizing $\hat H$. If the imaginary
part Im$(\xi_k (q))$ is positive for some $k$, the BEC in $\ket{0}$ is
dynamically unstable and pair production into the $\ket{\pm 1}$ state occurs.
By adapting the result of Ref.~\cite{Lamacraft2007} to the general case of spinor BECs in $F=2$ and $F=1$, we obtain three different regimes, which can be classified
according to the value of $q$. BECs in the $\ket{0}$ state are
(i) stable for $q>q_{cr}+|q_{cr}|$ (where $q_{cr}=-U_1n_0$); 
(ii) unstable for $q_{cr}<q<q_{cr}+|q_{cr}|$, where the instability rate of
the most unstable mode (with $k_{\rm max}=0$) is $\Lambda(q)=\sqrt{q_{cr}^2-(q-q_{cr})^2}/h$;
(iii) unstable for $q<q_{cr}$ with a constant instability rate $\Lambda(q)=|q_{cr}|/h$
for the most unstable modes $\hbar^2 k_{\rm max}^2/2m=q_{cr}-q$.
The $q$ dependence of $\Lambda(q)$ yields the magnetic field dependence of the pair creation efficiency.

Although the homogeneous picture offers important insights, the
observed $q$-dependence in trapped BECs is strikingly different, showing that 
the confinement must be considered to obtain even a qualitative understanding. 
This may be partially understood by comparison with the simple case of a $F=2$ BEC in a 
box potential $V(\vec r)=0$ for $r<R$, and $V(\vec r)=\infty$ otherwise. In the Thomas-Fermi 
regime the density $n_0$ of the BEC in the $\ket{0}$ state is approximately constant for $r<R$, and 
the energies of the Bogoliubov modes are $\xi_n^2(q)=(\epsilon_n+q)(\epsilon_n+q+2U_1n_0)$, with $\epsilon_n$ the 
energy of the $n$-th box state. Regimes (i) and (ii) are similar to the homogeneous case. 
However for regime (iii), occurring for $\eta(q)\equiv |q|-U_1 n_0>(\epsilon_0+\epsilon_1)/2$, 
the most unstable mode is given by the level $\bar{n}$ whose energy $\epsilon_{\bar n}$ is closest to $\eta(q)$. Contrary to the infinite system  
$\Lambda(q)=|\xi_{\bar n}(q)|/h$ shows pronounced maxima and minima, reaching its maximum $U_1n_0$ only 
when $\eta(q)$ is resonant with a box state.

For the actual experimental conditions not only the finite confinement but also 
the inhomogeneity of the potential and the related inhomogeneous Thomas-Fermi 
density profile $n_0(\vec r)$ contribute to the mode instability.
This is best understood by introducing the eigenfunctions of the effective Hamiltonian
$H_{eff}\phi_n(\vec r)=\epsilon_n\phi_n(\vec r)$. Note that the modes $\phi_n$ 
play a similar role to the box states in the simplified discussion above.
After projecting $\delta\hat \psi_{m_F}(\vec r)=\sum_n \phi_n(\vec r)\hat a_{n,m_F}$, 
we may re-write the Hamiltonian~(\ref{eq:H}) in the form:
\begin{eqnarray}
\hat H&=&\sum_{n,m_F=\pm 1}\left ( \epsilon_n +q \right ) \hat a_{n,m_F}^\dagger \hat a_{n,m_F}+ \nonumber \\
&&\sum_{n,n'}A_{n,n'} \left ( \hat a_{n,1}^\dagger \hat a_{n',-1}^\dagger + \hat a_{n',-1}\hat a_{n,1} \right )
\label{eq:He2}
\end{eqnarray}
where $A_{n,n'}=\int d^3{\vec r} \Omega_{eff}(\vec r)\phi_n(\vec
r)^*\phi_{n'}(\vec r)$. Note that the inhomogeneity of $V(\vec r)$ and $n_0(\vec r)$ 
plays a double role. First, $V_{eff}(\vec r)$ presents a non-trivial 
mexican-hat-like form (for $U_1>0$). Second,  
contrary to the simplified box potential, there is a significant coupling ($A_{n,n'}$) between 
different effective trap levels, induced by the inhomogeneity of $\Omega_{eff}(\vec r)$. 

After diagonalizing $\hat H$ we obtain the spin Bo\-go\-liu\-bov modes $\xi_\nu(q)$. Figure~\ref{fig2} (b) shows
the magnetic field dependence of the maximal instability rate 
$\Lambda(q)={\rm max}_\nu |{\rm Im}(\xi_\nu(q))|/h$ for
the experimental parameters of Fig.~\ref{fig2} (a). In the unstable regime
with low $|q|$ we may approximate the maximal instability rate
by $\Lambda(q)\simeq\sqrt{\tilde q_{cr}^2-(q-\tilde q_{cr})^2}/h$
with an effective $\tilde q_{cr}$ ($\simeq -30$~Hz in figure \ref{fig2}). However, as expected from the discussion above, this growth is not followed by a constant instability rate
for larger $|q|$. On the contrary, $\Lambda(q)$ shows pronounced maxima
and minima, which lead to a strongly enhanced or reduced pair creation efficiency
into $\ket{\pm 1}$ due to the exponential nature of the growth. Note that the position of the maxima and minima 
of the observed pair creation efficiency in our experiments is in excellent agreement with the 
position of the calculated maxima and minima of the instability rate in Fig. \ref{fig2}. 
However, the evaluation of the resonance strength is more subtle. The experimentally
observed resonance at low $|q|$ seems to be stronger than the one at high $|q|$, whereas the
calculated instability rate is slightly smaller for the resonance at low $|q|$. This is not
a disagreement, since the absolute population is not only determined by the growth rate
but also by the initial condition. In fact, the resonant growth can be initiated by spuriously
produced atoms in $\ket{\pm 1}$ and by vacuum spin fluctuations. Since these triggering mechanisms
contribute differently for the two resonances, they can fully explain the experimental results~\cite{KlemptNew}.

\begin{figure}[ht]
\centering
\includegraphics*[width=8.6cm]{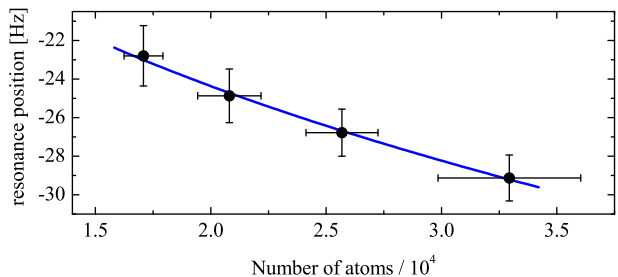}
\caption{Position of the lowest energy spinor dynamics resonance, recorded for various
atom numbers $N$ of the initial condensate in the $\ket{m_F=0}$ state. The blue line represents
a fit to a power law $N^\gamma$, yielding an exponent of $\gamma=0.36 \pm 0.02$.}
\label{fig3}
\end{figure}

The agreement in the resonance position is maintained for different experimental parameters. In particular, we 
vary the atom number in our experiments by introducing an additional hold time of up to $30$~ms
before the purification step, which allows for a controlled loss of atoms. By varying this hold time, the 
total atom number $N$ is changed without changing any of the other experimental
parameters. Figure \ref{fig3} shows the position of the resonance peak at low $|q|$ (obtained from a Gaussian fit) for different numbers of atoms. The resonant $q$ value increases with the number of atoms according to $N^{0.36 \pm 0.02}$.  
This exponent of $\approx 2/5$ closely resembles the dependence of the peak density on the number of atoms in the $\ket{0}$ state in the Thomas-Fermi regime. Hence the position of the resonance at low $|q|$ depends linearly on the BEC density, as expected from the linear scaling of $V_{eff}$ and $\Omega_{eff}$ with the density.

\begin{figure}[ht]
\centering
\includegraphics[width=8.6cm]{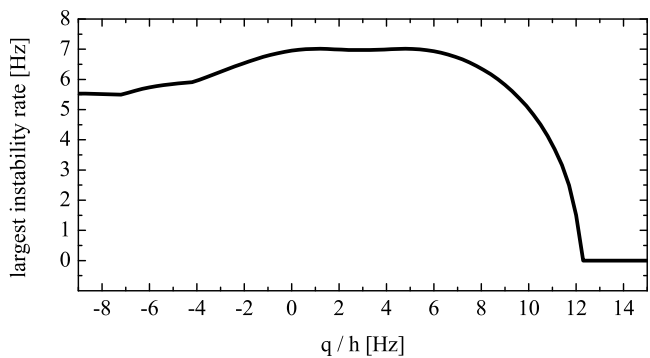}
\caption{Instability rate as a function of the quadratic Zeeman energy for the parameters of
Ref.~\cite{Leslie2008}, $(\omega_x,\omega_y,\omega_z)=2\pi\times(440,39,4.2)$~Hz and
$N=2
\times 10^6$ atoms. A constant maximal instability rate for $0<q<6$~Hz is
followed by a significant decay of the conversion efficiency for $q<0$.}
\label{fig4}
\end{figure}

Our results show that the pair-creation efficiency directly reflects the
instability of the spin Bogoliubov modes. This picture is
general and may be applied to other experiments on spinor BECs. In
particular, Fig.~\ref{fig4} shows the instability rate 
for the case of recent experiments with $^{87}$Rb
in $F=1$~\cite{Leslie2008}.
Our theory predicts a single resonance with a constant instability rate
for $0<q<6$~Hz,
and a clear decay of the pair-creation efficiency for $q<0$ (which was
unexpected from the
simplified homogeneous approximation), in excellent agreement with the
reported results~\cite{Leslie2008}.
We stress that the single-resonant character (as in
Ref.~\cite{Leslie2008})
or multi-resonant character (as in our case) of the dynamics largely
depends on atomic density and external confinement.

In conclusion, the pair creation efficiency in a spinor BEC initially in the $\ket{0}$ state
is characterized by a multi-resonant magnetic field dependence, 
which results from the non-trivial interplay between QZE, spin-changing collisions and
the finite size and inhomogeneity of both the trapping potential and the density of the BEC in the $\ket{0}$ state. 
We have shown that this inhomogeneity is crucial even for the qualitative understanding 
of the spinor dynamics. In particular, the trap-dependent modulation of the 
most unstable spin Bogoliubov mode is directly reflected by a strong enhancement or suppression of the 
spinor dynamics for particular magnetic field values. On the resonances, we observe
the spontaneous emergence of characteristic magnetization patterns~\footnote{Structure 
and symmetry of the patterns are described by the presented framework and will be published elsewhere.}.
It is also important to stress that, in particular for Rb BECs in $F=1$, the magnetic dipole-dipole
interaction may play a significant role not only in the spatial magnetization
pattern~\cite{Vengalattore2008,Swislocki2009,Sau2009}  but also in the pair production
efficiency. 

The characterization of the pair creation efficiency constitutes a comprehensive
description of a parametric amplifier for matter waves. The next challenge is the 
evaluation of the amplifier input which can be of classical or quantum nature. 
A realization where the presented parametric amplifier acts on pure vacuum spin 
fluctuations constitutes a promising source of non-classical states of matter~\cite{KlemptNew}.
This will open exciting perspectives for the detailed analysis of matter-wave
entanglement and squeezing during the parametric amplification of the $\ket{\pm 1}$
pairs~\cite{Duan2000,Pu2000}, possibly allowing for the production of atomic 
Einstein-Podolsky-Rosen pairs~\cite{Pu2000}.

\acknowledgments

We acknowledge support from the Centre for Quantum Engineering and Space-Time Research QUEST, from the
Deutsche Forschungsgemeinschaft (SFB 407), and the European Science Foundation (EuroQUASAR).

\bibliography{bibliothek}

\end{document}